\documentclass[final, authoryear, 5p, times, twocolumn]{elsarticle}

\usepackage{graphicx}
\usepackage{amssymb}
\usepackage[english]{babel}
\usepackage{lineno}
\usepackage{threeparttable}
\usepackage{tabularx}
\usepackage{color}
\usepackage[table]{xcolor}

\newcolumntype{L}[1]{>{\raggedright\arraybackslash}p{#1}}
\newcolumntype{C}[1]{>{\centering\arraybackslash}p{#1}}
\newcolumntype{R}[1]{>{\raggedleft\arraybackslash}p{#1}}

\begin{document}

\begin{frontmatter}

\title{A computer simulation protocol to assess the accuracy of a Radio Stereometric Analysis (RSA) image processor according to the ISO-5725}

\author[biomec]{Marco ~Bontempi\corref{cor1}}
\ead{marco.bontempi@ior.it}

\author[uniss]{Umberto ~Cardinale}
\author[unibo]{Laura ~Bragonzoni}
\author[clin2]{Giulio Maria ~Marcheggiani Muccioli}
\author[clin2]{Domenico ~Alesi}
\author[humanitas]{Berardo ~di Matteo}
\author[humanitas]{Maurilio ~Marcacci}
\author[clin2]{Stefano ~Zaffagnini}

\cortext[cor1]{Corresponding author. Tel.: +39 051 636 6852; fax: +39 051 583789}

\address[biomec]{SC Scienze e Tecnologie Chirurgiche, IRCCS Istituto Ortopedico Rizzoli, via di Barbiano 1/10, I-40136, Bologna, Italy}
\address[uniss]{Dipartimento di Scienze Mediche, Chirurgiche e Sperimentali, Universit\`a di Sassari, Viale San Pietro, 07100, Sassari (SS), Italy}
\address[unibo]{QUVI, Universit\`a di Bologna, corso d'Augusto 237, I-47921 Rimini (RN), Italy}
\address[clin2]{Clinica II, IRCCS Istituto Ortopedico Rizzoli, via Pupilli 1, I-40136, Bologna (BO), Italy}
\address[humanitas]{Istituto Clinico Humanitas, via Manzoni 56, I-20089, Rozzano (MI), Italy}

\date{\today}

%= ABSTRACT ================================================================================
\begin{abstract}
Radio-Sterometric-Analysis (RSA) and x-ray fluoroscopy require dedicated software to reconstruct the radiological scene and the position of the objects in space.
It is important to have a reliable validation to correctly use these softwares.
The two major regulations that deal with the definition of ``accuracy'' are the ISO-5725 and the GUM 1995.
The aim of this work, is to present a protocol for the evaluation of the accuracy of a radio stereometric software in terms of ``trueness'' and ``precision'', according to the standard ISO-5725.
The protocol consisted in a series of computer simulations of the radiological setup.
Each simulation changed the position and orientation of the x-ray sources, detectors and objects. 
Then, radiological images were generated.
The noise level of the images was also changed in order to evaluate the accuracy with different image qualities.
Then, the images can be processed with RSA softwares to evaluate their accuracy.
The protocol was tested on a custom RSA software developed at the Istituto Ortopedico Rizzoli and the accuracy of the results was evaluated.
The radiological scene reconstruction accuracy was found of the order of ($0.092 \pm 0.14$) mm for tube position and ($0.38 \pm 0.31$) mm/($2.09 \pm 1.39$) deg for detectors in the direction other than the source-detector direction.
In that case, in fact, the accuracy is of the order of ($2.68 \pm 3.08$) mm for the tube position and ($0.16 \pm 0.27$) mm/($0.75 \pm 1.16$) deg for detectors.
This fact is an intrinsic limitation of the scene reconstruction technique and it is widely discussed in the literature.
The model positioning and orientation evaluations was also very accurate: ($0.22 \pm 0.46$) mm/($0.26 \pm 0.22$) deg.
No differences were highlighted about the noise level. 
The accuracy remains the same independently from the the noise of the images.
The protocol was also useful to detect and fix hidden bugs in the software and optimize the algorithms.
\end{abstract}

%= KEYWORD =================================================================================
\begin{keyword}
accuracy \sep ISO-5725 \sep Radio-Stereometric-Analysis \sep RSA \sep image processing 
\end{keyword}

\end{frontmatter}

%\linenumbers

%= INTRODUCTION ============================================================================
\section{Introduction \label{sec:intro}}
The Radio-Stereometric-Analysis (RSA) is a technique to evaluate the micro-motions between bones and prosthesis \citep{Selvik1989, bruni2015}, or the modifications of the bones \citep{Callary2012, Martinkevich2015}.
The uniplanar fluoroscopy \citep{Catani2009} is used to evaluate the movements of bones and prosthesis during specific tasks \citep{Lawrence2018, Lin2013}.
Both these approaches require dedicated software to reconstruct the radiological scene and the position of the objects in space.
The problem of the validation of a new instrumentation or data processing algorithm is very common and thorny topic.
It is important to have a reliable validation to correctly use these softwares.
Currently, the standard method to evaluate the accuracy of a RSA image processing software is based on the work of \cite{Ranstam1999}.
In that work, the concept of accuracy was expressed in terms of upper limit of the 95\% of the confidence interval (CI) of the square root of the variance.
A regulation published by ISO was implicitly cited.

The framework of technical regulations includes the evaluation of the accuracy since the middle of the 1990s.
The two major regulations that face the definition of ``accuracy'' are the \cite{ISO5725} and the GUM 1995 and its updates \citep{GUM1995, BIPM2008}.
These two regulations start from different point of views: the ISO-5525 discusses the accuracy using the classical error analysis, while the GUM 1995 is focused on the concept of uncertainty.
By simplifying, both theoretically define the accuracy as the distance from the true value of the considered quantity.
But the GUM 1995 considers the accuracy as qualitative, because the true value is always unknown, while the ISO-5725 gives a quantitative definition.
According to the \cite{ISO5725-1}, the accuracy is the combination of ``trueness'' and ``precision''.
With these radical different philosophies it would be impossible to quantify the accuracy of a measurement system.
In this work we would like to propose the idea that the accuracy can be quantitatively evaluated according to ISO definition, without contradicting the GUM definition: the true value can be provided by a computer simulation.
In this case, in fact, the values of the measurable quantities are decided by the software and they can be considered as the true value.

The aim of this work is to present a protocol for the evaluation of the accuracy of a RSA and fluoroscopy softwares.
Then we want to use the protocol to evaluate the accuracy of the output data processed with the custom software developed at the Istituto Ortopedico Rizzoli (IOR), Bologna, Italy.
The software reconstructs the radiological scene, and the position of a model in space according to the acquired images (Model-Based RSA, \cite{kaptein2003}).
The \cite{ISO16087} defined the accuracy of an RSA examination.
The mentioned documents referred to the works of \cite{Ornstein2000} and \cite{Bragdon2002}, and are based on experimental results of a real setup.
On the contrary, in this work the interest is the evaluation of the accuracy of the software used to evaluate the data.

Our hope is to provide a useful tool that can standardize the evaluation of the accuracy and allow an easier comparison between different data processing systems.

%= MATERIAL AND METHODS ====================================================================
\section{Materials and Methods\label{sec:MandM}}
A simulation software was developed to evaluate the accuracy of radio metric (uniplanar and biplanar) data processing softwares.

\subsection{The simulation software}
\begin{figure}[h]
	\centering
	\includegraphics[width=8.6cm]{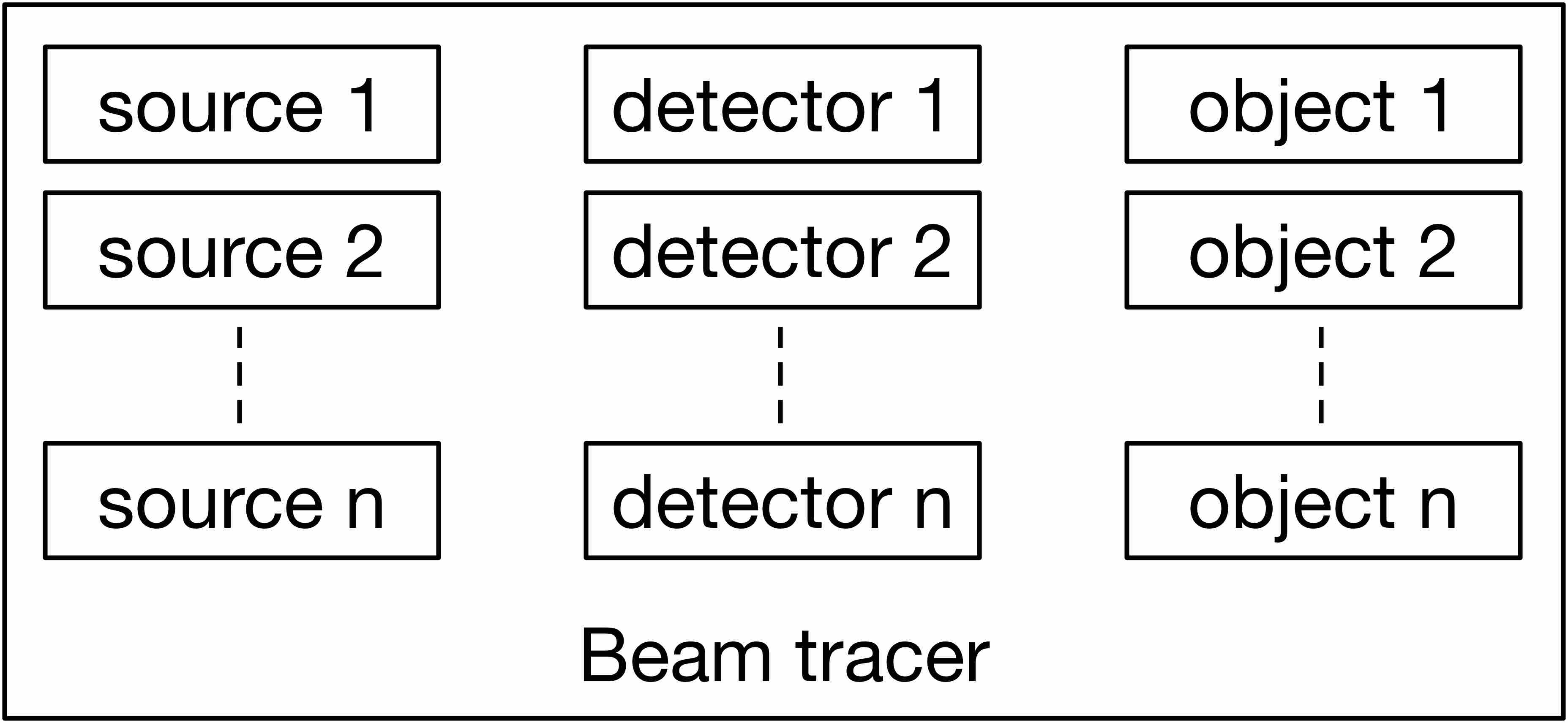}
	\caption{\label{fig:chart} Flow chart of the simulation software used to test the radio stereometric software.}
\end{figure}

The simulation software was composed of 4 parts: x-ray source simulator, x-ray detector simulator, object in space simulator, and the x-ray beam tracer.
The software received as inputs the position and orientation of the x-ray sources, detectors and objects.
Then the output was a set of images that represents the objects according to the setup.
The figure \ref{fig:chart} represents the structure of the described software.
%~ chart ~~~~~~~~~~~~~~~~~~~~~~~~~~~~~~~~~~~~~~~~~~~~~~~~~~~~~~~~~~~~~~~~~~~~~~~~~~~~~~~~~~~
%~~~~~~~~~~~~~~~~~~~~~~~~~~~~~~~~~~~~~~~~~~~~~~~~~~~~~~~~~~~~~~~~~~~~~~~~~~~~~~~~~~~~~~~~~~~
Each simulated object had its own reference system $(\vec{p}, R)$, where $\vec{p}$ was the position and $R$ was a $3\times3$ matrix representing the orientation.

The x-ray tube was simulated as an ideal monocromatic point-like source.
It was characterized only by its position in space and had the scope to be the starting point of the x-ray beam.

The detector was a matrix of the end points of the x-ray beam. 
Its input parameters were the bits per pixel of the image (bpp), the number of pixels in the matrix and the reference system to determine its position and orientation in space. 
The beam tracer scanned the matrix, and for each matrix element, calculated the ending point of the x-ray beam and, using the x-ray source position, defined the beam path.

The object in space was a phantom whose input was the position/orientation parameters.
The material was not important. 
The beam tracer calculated if the beam path crossed the object.
If yes, the value $I_{max}$ was assigned to the corresponding pixel.
The $I_{max}$ was calculated from the bbp assigned to the detector and it was $I_{max} = 2^{bpp} - 1$.
Because the theme of this work is RSA, the phantom can be a set of markers, or a bone model, or a prosthesis model according to the different tested methodology.
The user choses the proper phantom.

Once the scanning of the detector was completed and the image of the object was created, the simulator added the noise to the image.
The noise average level was added to change the contrast of the image and analyzes the performances of the software with different image conditions.
The noise level was set by the user as percentage of the $I_{max}$.
Thus, using the set noise level as average parameter, a random Poisson value was added to all the pixel of the image that did not contain the model.

\subsection{The accuracy evaluation protocol}
The core of the accuracy defined by the ISO-5725 is the evaluation of trueness and precision.
The first is the distance from the true value, while the second is the spread (repeatability) of the data.
These two quantities have to be evaluated both for the radiological scene reconstruction and for the object position.
To evaluate these quantities, a simulation evaluation protocol was set according to the following steps:
\begin{enumerate}
	\item Simulation setup: number of sources detectors, calibration tools, objects, and image sizes
	\item Simulation run: 5 set of 20 simulation runs were performed
	\item Each set of simulations changed the noise level from 0.0\% (ideal image), to 90\% (high noise level), with increment steps of 22.5\% in order to have 5 set of images. Each run posed the sources and the detectors at random positions and orientations
	\item Each simulation returned 2 set of data: the first was the scene calibration, the second was the image of the projected object
	\item From the calibration data, the trueness and precision of the radiological scene reconstruction were evaluated
	\item From the object images, the trueness and precision of the object position was calculated
\end{enumerate}
The step 5 and 6 of this procedure were let generic on purpose, because the calibration of the scene changed case-by-case.
The steps from 1 to 4 were handled by the simulator, while the steps 5 and 6 were handled by the image processing software under test.
The choice of changing the setup and the object positions and orientations at every run were made because presenting always the same simulation to the software and to the operator could trigger some automatism that affects the correctness of the results.
Because the core of the accuracy evaluations was the differences between the set and obtained values, these can be calculated at every run and are independent from the setup of the simulation.

The trueness was defined as the mean of the residuals ($\Delta$) between the true values ($t_{xi}$), set by the simulator, and the obtained values ($x_{i}$) calculated by the data software (equation \ref{eq:trueness}).
%. trueness ................................................................................
\begin{equation}
	\label{eq:trueness}
	T_{x} = \frac{1}{N} \sum_{i = 1}^{N} \Delta(x_{i}, t_{i})
\end{equation}
%...........................................................................................
According to the \cite{ISO5725-3}, the precision was calculated as the 95\% of confidence interval of the standard deviation of the mean (equation \ref{eq:precision}) of the residuals $\Delta(x, t)$.
%. precision ...............................................................................
\begin{equation}
	\label{eq:precision}
	P_{x} = \frac{\tau}{\sqrt{N}} \sqrt{\frac{1}{N - 1}\sum_{i = 1}^{N}\left( \Delta(x_{i}, t_{i}) - T_{x} \right)^{2}} 
\end{equation}
%...........................................................................................
where the symbol $x$ indicates one of the position/orientation parameters.
The notation $\Delta$ was used because the difference depends on the parameter type: position or orientation.
The constant $\tau$ is the \emph{t-value} calculated on N random variables ($\Delta$) normally distributed.
In this case, for 20 simulations, its value was 1.73.
These equations represent the mean distance with sign from the true value.
This is because it is important to know if the bias is greater or lesser than the true value.
A root mean distance, as used by \cite{Ranstam1999}, lose this information. 

The position accuracy can be evaluated by comparing the coordinates of the obtained and of the true values setting $\Delta(x, t) = (x - t)$.

A more complex problem was the evaluation of the accuracy of the orientation.
In this case, two orthonormal matrices have to be compared.
Because it is easier, and more clear, to work with numbers, the matrix differences have to be transposed in representative numbers.
The procedure started from the calculation of the transformation matrix ($M$) between the reconstructed ($R_{rec}$) and the true matrix ($R_{true}$) (equation \ref{eq:trans}).
%. trans ...................................................................................
\begin{equation}
	\label{eq:trans}
	M = R_{rec} R_{true}^{-1}
\end{equation}
%...........................................................................................
This matrix $M$ represents the difference between the true and the evaluated orientation.
Then the Eulerian decomposition of $M$ was calculated, according to the 3 axes:
%. trans ...................................................................................
\begin{equation}
	\label{eq:trans}
	M = M_{x}(\alpha) M_{y}(\beta) M_{z}(\gamma)
\end{equation}
%...........................................................................................
In this way, the 3 angles, $\alpha$, $\beta$ and $\gamma$ represented the $\Delta$ around the reference axes.

\begin{figure}[h]
	\centering
	\includegraphics[width=8.6cm]{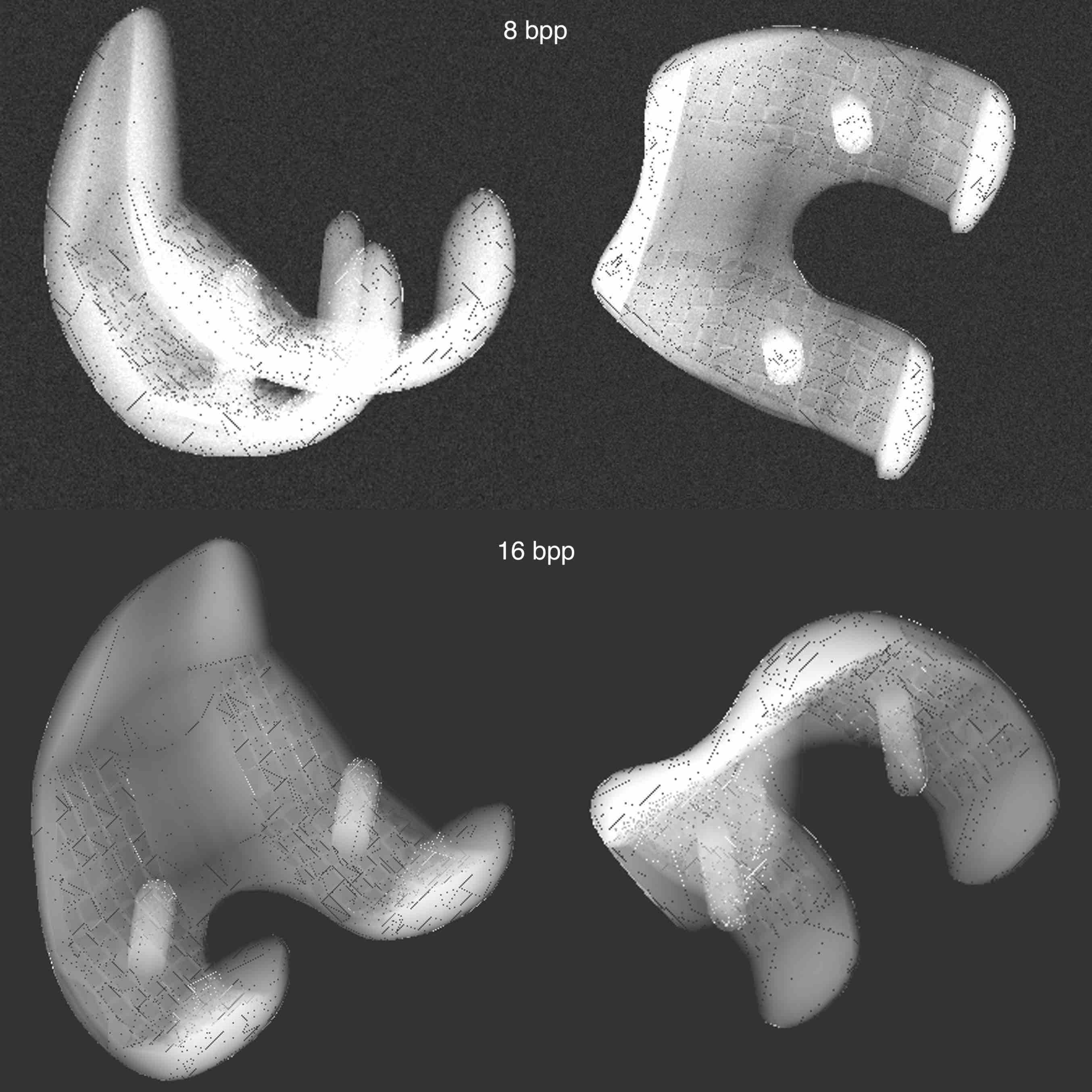}
	\caption{\label{fig:ref} Schema of the RSA setup with the direction used to show the results of the scene reconstruction. The background show the real RSA device developed at IOR (Bi-Stand DRX, ASSING Group, Rome, Italy). The overlapped paint represents the simulated setup.}
\end{figure}

Concerning the presentation of the results, a special notation was used.
Because the two beamlines had different orientations (figure \ref{fig:ref}), the presentation of the results according to the coordinates provided by the calibration box could generate confusion.
%~ ref ~~~~~~~~~~~~~~~~~~~~~~~~~~~~~~~~~~~~~~~~~~~~~~~~~~~~~~~~~~~~~~~~~~~~~~~~~~~~~~~~~~~~~
%~~~~~~~~~~~~~~~~~~~~~~~~~~~~~~~~~~~~~~~~~~~~~~~~~~~~~~~~~~~~~~~~~~~~~~~~~~~~~~~~~~~~~~~~~~~
For this reason, to make the results of the scene reconstruction more clear, subscript labels were added to the spatial and angular accuracies according to three representative direction: detector rows direction (subscript label ``r''), detector columns direction (subscript label ``c'') and source-detector direction (subscript label ``s'').
In this way it should be easier to compare the results.
Instead, the accuracies of the position/orientation of the model were presented without any specific subscript label.

\subsection{Validation}
\begin{figure}[h]
	\centering
	\includegraphics[width=8.6cm]{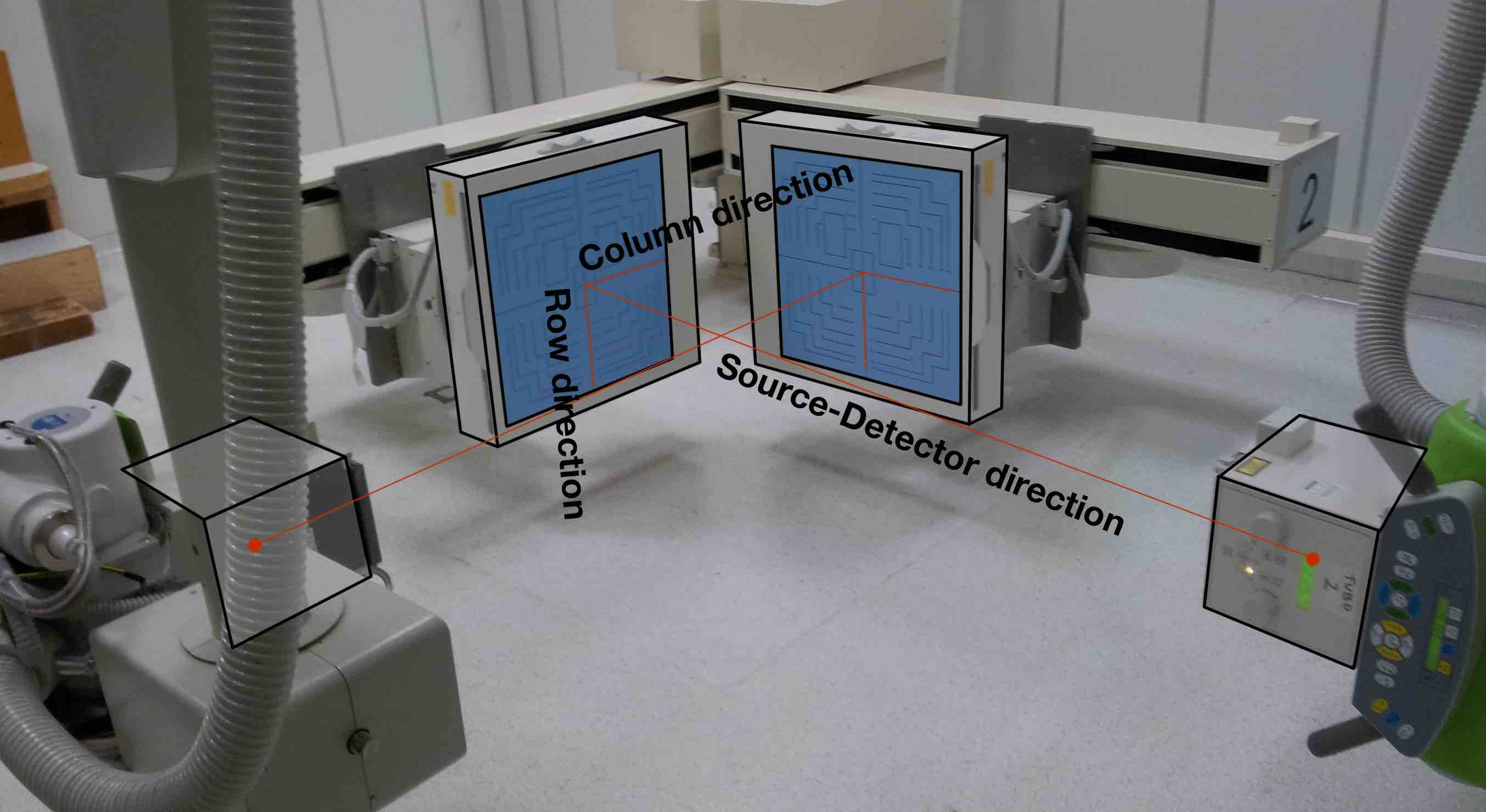}
	\caption{\label{fig:reference} Reference system defined in the simulated setup.}
\end{figure}

The described protocol was used to test the custom RSA software developed at IOR.
The simulated scene was composed by two sources and two detectors that make two orthogonal beamlines.
The detector dimensions were $43\times43$ cm with a pixel matrix of $1440\times1440$ pixels.
Each beamline had the source-to-detector distance set to 180 cm.
The coordinates reference system assigned to the radiological setup is showed in table \ref{fig:reference}.
The beamline 1 lies along the Z axis, while the beamline 2 lies along the X axis. 
The arrows show the positive rotation direction of the Euler angles determined for the orientations.

%~ reference ~~~~~~~~~~~~~~~~~~~~~~~~~~~~~~~~~~~~~~~~~~~~~~~~~~~~~~~~~~~~~~~~~~~~~~~~~~~~~~~
%~~~~~~~~~~~~~~~~~~~~~~~~~~~~~~~~~~~~~~~~~~~~~~~~~~~~~~~~~~~~~~~~~~~~~~~~~~~~~~~~~~~~~~~~~~~

When the radiological scene was defined, the RSA protocol was simulated.
%~ objects ~~~~~~~~~~~~~~~~~~~~~~~~~~~~~~~~~~~~~~~~~~~~~~~~~~~~~~~~~~~~~~~~~~~~~~~~~~~~~~~~~
\begin{figure}[h]
	\centering
	\includegraphics[width=8.6cm]{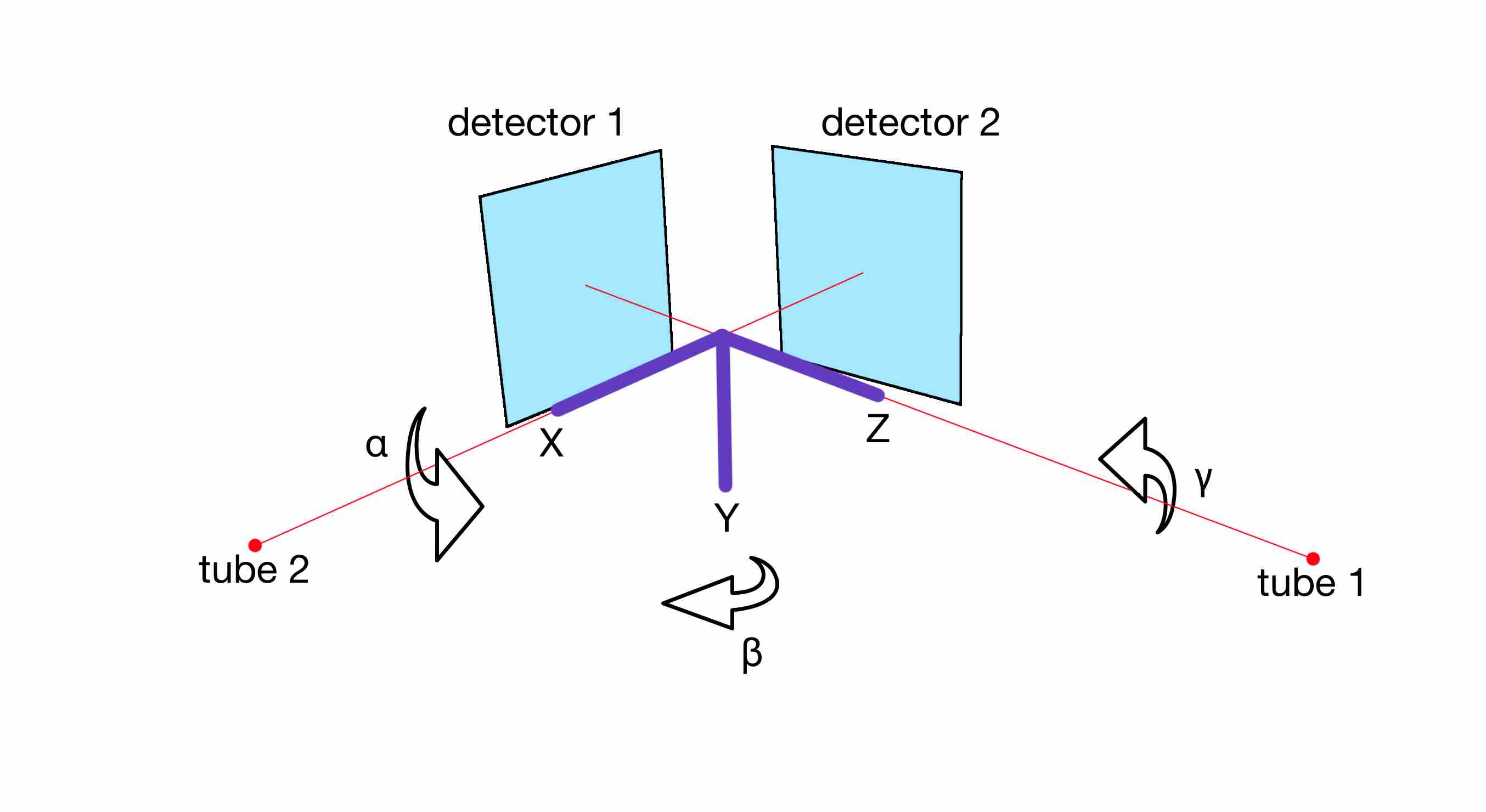}
	\caption{\label{fig:objects} Figure of the objects used in the simulations. On the left, the calibration box with the markers used to reconstruct the radiological scene. On the right the model used to test the model position accuracy. The sizes of the two objects are not in scale.}
\end{figure}
%~~~~~~~~~~~~~~~~~~~~~~~~~~~~~~~~~~~~~~~~~~~~~~~~~~~~~~~~~~~~~~~~~~~~~~~~~~~~~~~~~~~~~~~~~~~
The aim was the acquisition of the simulated images of the calibration box and of a model  (figure \ref{fig:objects}).
The model of a femoral knee prosthesis was chosen (figure \ref{fig:objects} on the right). 
It was randomly posed in space at each run of the simulator, changing its position and orientation.
At each run, the images of the calibration box and of the prosthesis were acquired and processed using the RSA software as if they came from a real RSA examination.

The tested software was written in MATLAB\textsuperscript{\textregistered} (R2017a, The MathWorks Inc, Natick, MA, USA).
The software acquired the images of the calibration box and of the prosthetic model.
Then it used images of the calibration box to reconstruct the position of the sources and detectors.
The model position was reconstructed in two steps.
The first step consisted of the segmentation of the images acquired by the simulator.
Then a cad model of the segmented object was posed in space to fit the edges extracted from the images \citep{Kaptein2004}.

Because the software had semi-automatic algorithms, an expert human operator analyzed all the images as they were generated by a common RSA exam.
Then, according to the described protocol, the radiological scene and the model position was evaluated for each run of the simulator and  the trueness and the precision were evaluated.

%= RESULTS =================================================================================
\section{Results}
\begin{figure}[h]
	\centering
	\includegraphics[width=8.6cm]{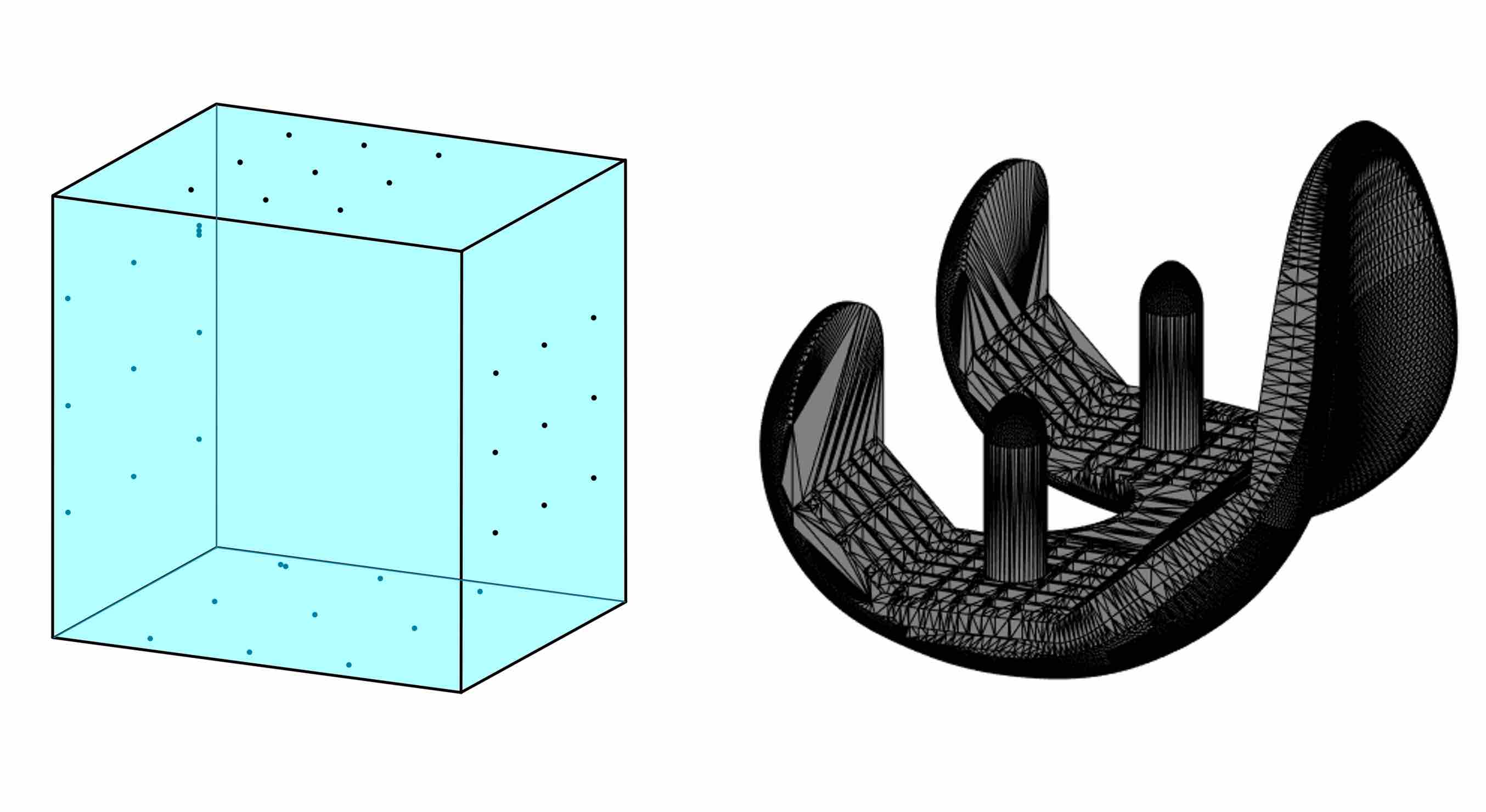}
	\caption{\label{fig:sim} Example of images generated by the simulator. Each row shows calibration phantom (left) and the model (right) at different noise levels.}
\end{figure}

The results of the accuracy evaluation are showed in the following table.
By convention, all the results are showed with two decimal digits.
An example of the images generated by the simulator is showed in figure \ref{fig:sim}.
%~ sim ~~~~~~~~~~~~~~~~~~~~~~~~~~~~~~~~~~~~~~~~~~~~~~~~~~~~~~~~~~~~~~~~~~~~~~~~~~~~~~~~~~~~~
%~~~~~~~~~~~~~~~~~~~~~~~~~~~~~~~~~~~~~~~~~~~~~~~~~~~~~~~~~~~~~~~~~~~~~~~~~~~~~~~~~~~~~~~~~~~

The first results are the accuracy of the radiological scene reconstruction, \emph{i.e.} the position and orientation of x-ray sources and detectors, as a function of the image contrast of the calibration box.
The results of the accuracy evaluation of the x-ray tube positions are showed in table \ref{tab:tubes}.
%~ tubes ~~~~~~~~~~~~~~~~~~~~~~~~~~~~~~~~~~~~~~~~~~~~~~~~~~~~~~~~~~~~~~~~~~~~~~~~~~~~~~~~~~~
\begin{table*}[h]
	\centering
	\begin{tabular}{r|ccc|ccc}
	& \multicolumn{3}{c|}{tube 1} & \multicolumn{3}{c}{tube 2}\\
	Noise & $x_{r}$ (mm) & $y_{c}$ (mm) & $z_{s}$ (mm) & $x_{s}$ (mm) & $y_{r}$ (mm) & $z_{c}$ (mm)\\
	\hline\hline
	   0\% & $-0.036 \pm 0.093$ & $0.020 \pm 0.22$ & $-1.11 \pm 2.46$ & \cellcolor{blue!25}$-6.22 \pm 5.47$ & $0.054 \pm 0.23$ & $-0.020 \pm 0.022$ \\
	22.5\% &  $0.0069 \pm 0.081$ & $0.083 \pm 0.16$ & $-0.28 \pm 1.93$ & $-3.84 \pm 4.03$ & $0.12 \pm 0.18$ & $-0.091 \pm 0.17$  \\ 
	45.0\% & $0.027 \pm 0.085$ & $0.075 \pm 0.14$ & $-0.31 \pm 1.66$ & $-2.82 \pm 3.55$ & $0.098 \pm 0.15$ & $-0.13 \pm 0.15$  \\ 
	67.5\% & $0.034 \pm 0.077$ & $0.074 \pm 0.12$ & $-0.66 \pm 1.58$ & $-2.25 \pm 3.16$ & $0.12 \pm 0.15$ & \cellcolor{blue!25}$-0.16 \pm 0.14$  \\ 
	90.0\% &  $0.055 \pm 0.077$ & $0.079 \pm 0.12$ & $-0.26 \pm 1.57$ & $-1.87 \pm 2.97$ & $0.13 \pm 0.14$ & \cellcolor{blue!25}$-0.17 \pm 0.12$  \\
	\hline\hline
	\end{tabular}
	\caption{\label{tab:tubes} (trueness $\pm$ precision) of the position of the x-ray sources used in the simulations.}
\end{table*}
%~~~~~~~~~~~~~~~~~~~~~~~~~~~~~~~~~~~~~~~~~~~~~~~~~~~~~~~~~~~~~~~~~~~~~~~~~~~~~~~~~~~~~~~~~~~
The table lists the (trueness $\pm$ precision) of the positions of the two simulated x-ray sources as a function of the image noise level.
The orientation of the sources was not calculated, because they were considered as pointlike.

\begin{figure}[h]
	\centering
	\includegraphics[width=8.6cm]{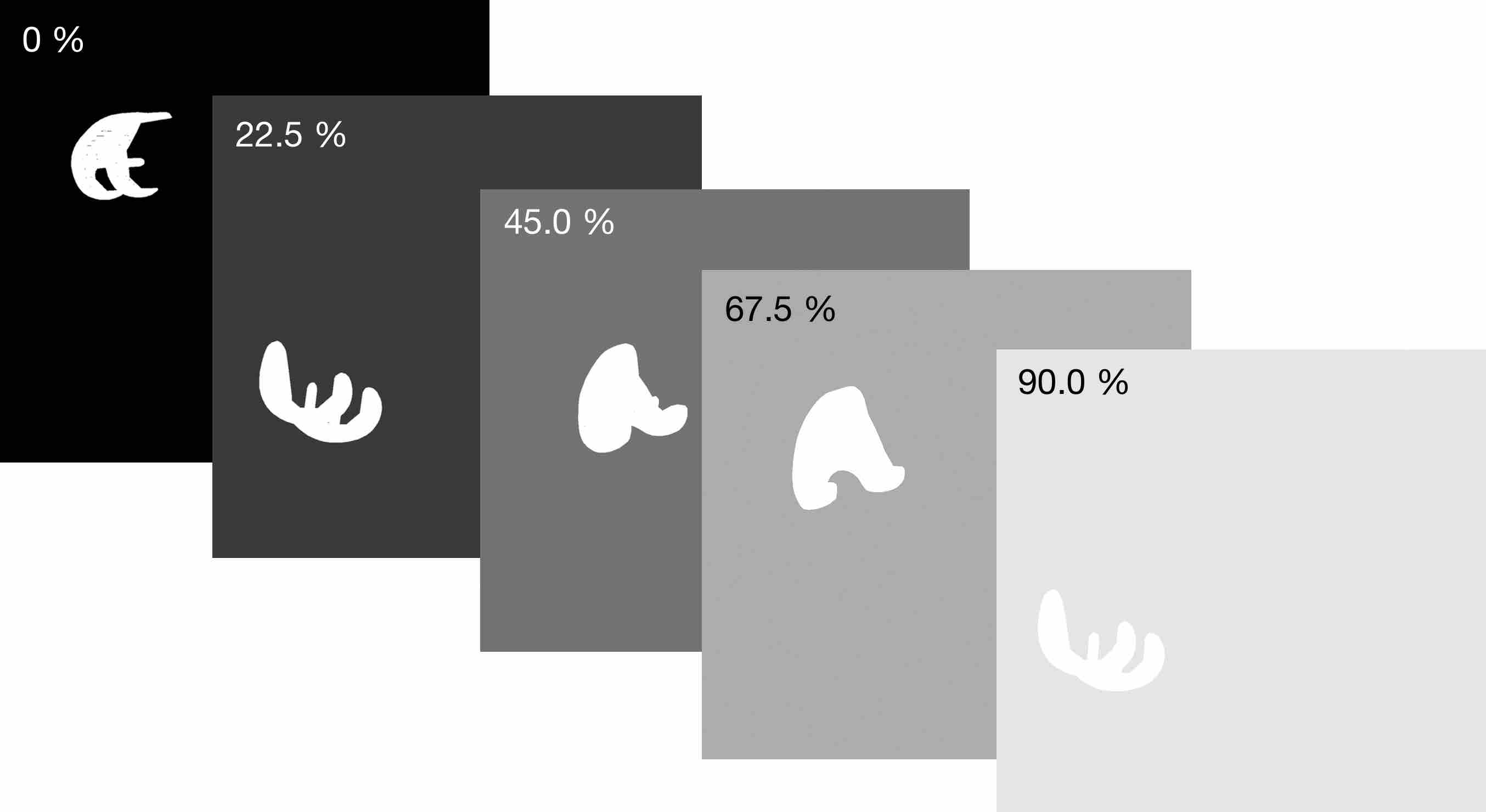}
	\caption{\label{fig:scene} Demonstrative picture of the comparison between the set and reconstructed beamlines. The red objects are the true objects, while the blue are the obtained from the image processing software. The circles show the different reconstruction accuracy of the x-ray source position for beamline 1 and 2. The beamlines appear inverted with respect to the figure \ref{fig:reference}, because the reference system was oriented with the Y axis in the upper direction.}
\end{figure}

The detector position/orientation accuracies are showed in tables \ref{tab:dect1} and \ref{tab:dect2}.
%~ dect1 ~~~~~~~~~~~~~~~~~~~~~~~~~~~~~~~~~~~~~~~~~~~~~~~~~~~~~~~~~~~~~~~~~~~~~~~~~~~~~~~~~~~
\begin{table*}[h]
	\centering
	\begin{tabular}{r|ccc|ccc}
	    	     & \multicolumn{3}{c|}{orientation} & \multicolumn{3}{c}{position} \\
	Noise & $\alpha_{r}$ (deg) & $\beta_{c}$ (deg) & $\gamma_{s}$ (deg) & $x_{r}$ (mm) & $y_{c}$ (mm) & $z_{s}$ (mm) \\
		\hline\hline
	   	   0\%.& $-0.063 \pm 0.62$ & \cellcolor{red!25}$0.59 \pm 0.27$ & \cellcolor{red!25}$0.055 \pm 0.017$ & \cellcolor{red!25}$0.82 \pm 0.18$  & \cellcolor{blue!25}$-0.33 \pm 0.092$ & $0.71 \pm 2.089$ \\ 
		22.5\% & $0.0051 \pm 0.46$ & \cellcolor{red!25}$0.45 \pm 0.24$ & \cellcolor{red!25}$0.037 \pm 0.015$ & \cellcolor{red!25}$0.82 \pm 0.13$  & \cellcolor{blue!25}$-0.24 \pm 0.085$ & $1.18 \pm 1.78$  \\
		45.0\% & $0.055 \pm 0.39$ & \cellcolor{red!25}$0.30 \pm 0.23$ & \cellcolor{red!25}$0.034 \pm 0.013$ & \cellcolor{red!25}$0.81 \pm 0.11$  & \cellcolor{blue!25}$-0.23 \pm 0.081$ & $0.79 \pm 1.52$  \\ 
		67.5\% & $0.068 \pm 0.36$ & \cellcolor{red!25}$0.25 \pm 0.21$ & \cellcolor{red!25}$0.034 \pm 0.012$ & \cellcolor{red!25}$0.80 \pm 0.10$  & \cellcolor{blue!25}$-0.23 \pm 0.073$ & $0.53 \pm 1.51$  \\ 
		90.0\% & $0.028 \pm 0.34$ & $0.18 \pm 0.22$ & \cellcolor{red!25}$0.032 \pm 0.011$ & \cellcolor{red!25}$0.81 \pm 0.092$ & \cellcolor{blue!25}$-0.21 \pm 0.073$ & $0.32 \pm 1.57$  \\
		\hline\hline
	\end{tabular}
	\caption{\label{tab:dect1} (trueness $\pm$ precision) of detector reconstruction accuracy in beamline 1, as a function of the image contrast.}
\end{table*}
%~~~~~~~~~~~~~~~~~~~~~~~~~~~~~~~~~~~~~~~~~~~~~~~~~~~~~~~~~~~~~~~~~~~~~~~~~~~~~~~~~~~~~~~~~~~
%~ dect2 ~~~~~~~~~~~~~~~~~~~~~~~~~~~~~~~~~~~~~~~~~~~~~~~~~~~~~~~~~~~~~~~~~~~~~~~~~~~~~~~~~~~
\begin{table*}[h]
	\centering
	\begin{tabular}{r|ccc|ccc}
		 & \multicolumn{3}{c|}{orientation} & \multicolumn{3}{c}{position} \\
	Noise & $\alpha_{s}$ (deg) & $\beta_{r}$ (deg) & $\gamma_{c}$ (deg) & $x_{s}$ (mm) & $y_{r}$ (mm) & $z_{c}$ (mm) \\
		\hline\hline
		   0\% & $-0.12 \pm 0.49$ & \cellcolor{blue!25}$-1.15 \pm 0.36$ & \cellcolor{red!25}$-0.020 \pm 0.016$ & \cellcolor{red!25}$0.79 \pm 0.25$ & \cellcolor{blue!25}$-0.29 \pm 0.13$ & \cellcolor{blue!25}$-7 29 \pm 3.47$ \\
		22.5\% & $-0.18 \pm 0.40$ & \cellcolor{blue!25}$-0.47 \pm 0.41$ & \cellcolor{blue!25}$-0.024 \pm 0.013$ & \cellcolor{red!25}$0.79 \pm 0.18$ & \cellcolor{blue!25}$-0.21 \pm 0.11$ & \cellcolor{blue!25}$-3.94 \pm 2.97$ \\
		45.0\% & $-0.27 \pm 0.36$ & $-0.20 \pm 0.40$ & \cellcolor{blue!25}$-0.024 \pm 0.011$ & \cellcolor{red!25}$0.81 \pm 0.15$ & \cellcolor{blue!25}$-0.17 \pm 0.10$ & \cellcolor{blue!25}$-2.75 \pm 2.60$ \\
		67.5\% & $-0.26 \pm 0.33$ & $-0.12 \pm 0.36$ & \cellcolor{blue!25}$-0.025 \pm 0.011$ & \cellcolor{red!25}$0.79 \pm 0.14$ & \cellcolor{blue!25}$-0.16 \pm 0.089$ & $-2.15 \pm 2.36$ \\
		90.0\% & $-0.23 \pm 0.33$ & \cellcolor{blue!25}$-0.77 \pm 0.34$ & \cellcolor{blue!25}$-0.022 \pm 0.010$ & \cellcolor{red!25}$0.78 \pm 0.13$ & \cellcolor{blue!25}$-0.17 \pm 0.085$ & $-1.76 \pm 2.31$ \\
		\hline\hline
	\end{tabular}
	\caption{\label{tab:dect2} (trueness $\pm$ precision) of detector reconstruction accuracy in beamline 2, as a function of the image contrast.}
\end{table*}
%~~~~~~~~~~~~~~~~~~~~~~~~~~~~~~~~~~~~~~~~~~~~~~~~~~~~~~~~~~~~~~~~~~~~~~~~~~~~~~~~~~~~~~~~~~~
The figure \ref{fig:scene} shows an example of the comparison between the set and the reconstructed position/orientation of a beamline.
%~ scene ~~~~~~~~~~~~~~~~~~~~~~~~~~~~~~~~~~~~~~~~~~~~~~~~~~~~~~~~~~~~~~~~~~~~~~~~~~~~~~~~~~~
%~~~~~~~~~~~~~~~~~~~~~~~~~~~~~~~~~~~~~~~~~~~~~~~~~~~~~~~~~~~~~~~~~~~~~~~~~~~~~~~~~~~~~~~~~~~

\begin{figure}[h]
	\centering
	\includegraphics[width=8.6cm]{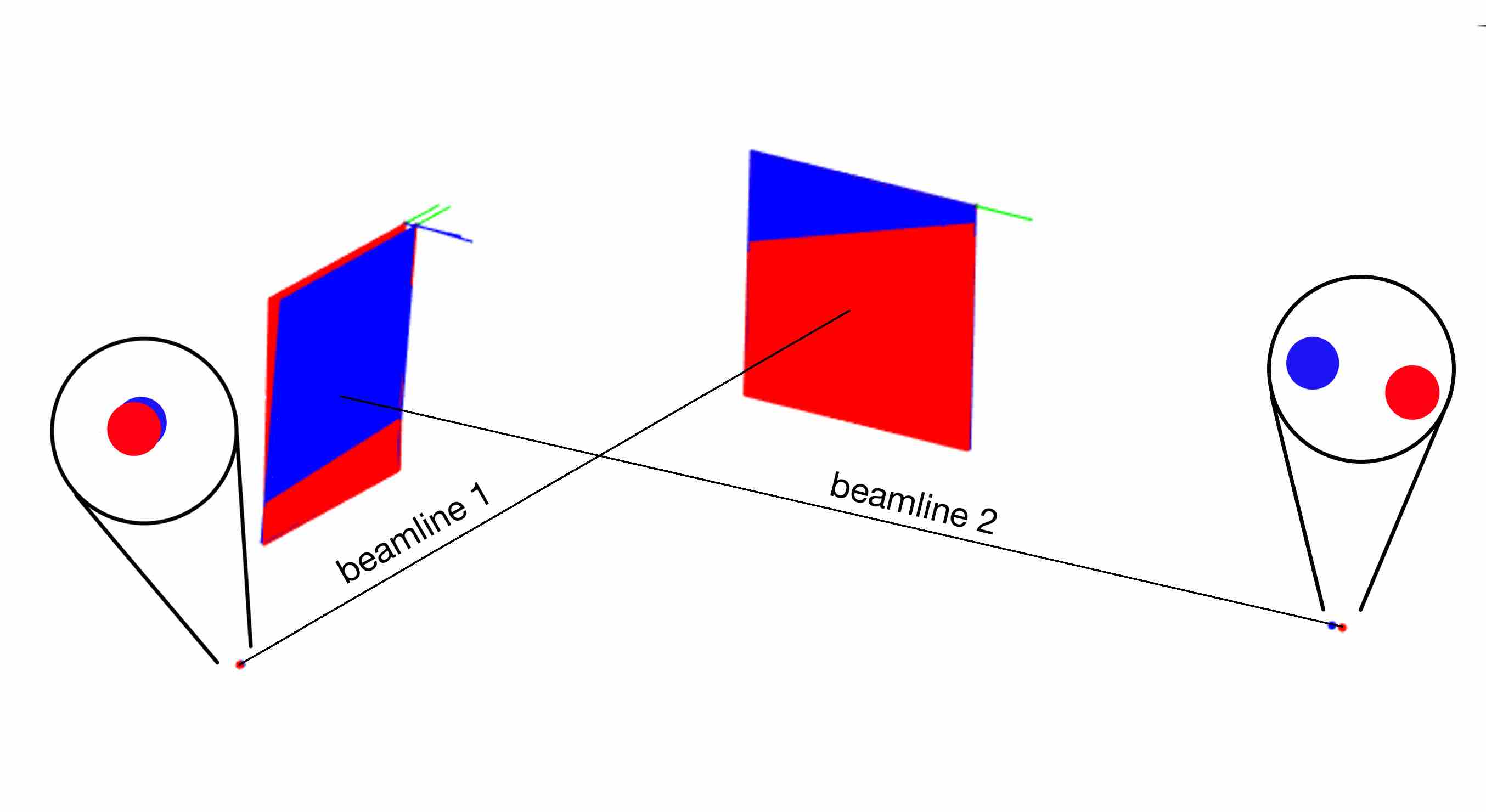}
	\caption{\label{fig:prmodel} Dimostrative picture of the comparison between the set and reconstructed model position. The red model is the true object, while the blue is the obtained from the image processing software.}
\end{figure}

The model position/orientation results are showed in table \ref{tab:model}.
%~ model ~~~~~~~~~~~~~~~~~~~~~~~~~~~~~~~~~~~~~~~~~~~~~~~~~~~~~~~~~~~~~~~~~~~~~~~~~~~~~~~~~~~
\begin{table*}[h]
	\centering
	\begin{tabular}{r|ccc|ccc}
	    & \multicolumn{3}{c|}{orientation} & \multicolumn{3}{c}{position} \\
	Noise & $\alpha$ (deg) & $\beta$ (deg) & $\gamma$ (deg) & $x$ (mm) & $y$ (mm) & $z$ (mm) \\
		\hline\hline
		   0\% & \cellcolor{blue!25}$-0.40 \pm 0.31$ & $0.012 \pm 0.51$ & $-0.16 \pm 0.27$ &  $0.017 \pm 0.22$ & \cellcolor{red!25}$0.55 \pm 0.16$ & \cellcolor{blue!25}$-0.19 \pm 0.18$ \\
		22.5\% & $-0.18 \pm 0.33$ & $-0.26 \pm 0.70$ & $0.034 \pm 0.26$ & $0.060 \pm 0.26$ & \cellcolor{red!25}$0.46 \pm 0.15$ & \cellcolor{blue!25}$-0.26 \pm 0.22$ \\
		45.0\% & $-0.24 \pm 0.36$ & $-0.24 \pm 0.62$ & $0.055 \pm 0.26$ & $0.056 \pm 0.23$ & \cellcolor{red!25}$0.33 \pm 0.18$ & $-0.21 \pm 0.24$ \\
		67.5\% & $-0.25 \pm 0.44$ & $-0.35 \pm 0.73$ & $0.038 \pm 0.26$ & $0.026 \pm 0.26$ & \cellcolor{red!25}$0.31 \pm 0.19$ & $-0.23 \pm 0.25$ \\
		90.0\% & $-0.26 \pm 0.34$ & $-0.26 \pm 0.68$ & $0.024 \pm 0.25$ & $-0.0089 \pm 0.24$ & \cellcolor{red!25}$0.30 \pm 0.18$ & $-0.16 \pm 0.24$ \\
		\hline\hline
	\end{tabular}
	\caption{\label{tab:model} (trueness $\pm$ precision) of the model position/orientation accuracy.}
\end{table*}
%~~~~~~~~~~~~~~~~~~~~~~~~~~~~~~~~~~~~~~~~~~~~~~~~~~~~~~~~~~~~~~~~~~~~~~~~~~~~~~~~~~~~~~~~~~~
The table shows the values of accuracy of the considered 6 parameters.
Figure \ref{fig:prmodel} shows a comparison between the true position of the model and the one evaluated with the software.
%~ prmodel ~~~~~~~~~~~~~~~~~~~~~~~~~~~~~~~~~~~~~~~~~~~~~~~~~~~~~~~~~~~~~~~~~~~~~~~~~~~~~~~~~
%~~~~~~~~~~~~~~~~~~~~~~~~~~~~~~~~~~~~~~~~~~~~~~~~~~~~~~~~~~~~~~~~~~~~~~~~~~~~~~~~~~~~~~~~~~~

%= DISCUSSION ==============================================================================
\section{Discussion}
The proposed protocol has the aim to evaluate the accuracy of a image processing software dedicated to RSA, or in general to radio metric softwares. 

The results of the protocol are compliant with the ISO-5725 in order to have a standard reference and compare the performances of different softwares and results.
The software accuracy was divided in scene reconstruction and model positioning.
The simulation of different levels of image noise gives the opportunity to explore the behaviour of the algorithms in ideal conditions (low noise) and in critical conditions (high noise).
This could allow developers to optimize their codes, or to discover unwanted and hidden bugs.

Concerning the scene reconstruction, the table \ref{tab:tubes} shows that the tube position evaluation has a sub-millimetric accuracy in the directions different from the source-detector direction.
In the source-detector direction, the accuracy is more rough and uses the order of millimeters for the trueness and centimeters for the precision.
This difference is also reported in the literature \citep{Tsai1987, Hartley2004} and it is an intrinsic defect of the setup.
An other difference that have to be highlighted is the different accuracies of the source positions in the two beamlines, in the source-detector direction.
Looking at the table \ref{tab:tubes}, it is clear that the values in the columns labelled with ``s'' are quite different. 
The second beamline has values more than doubled with respect to the first.
Moreover, the sign of the reconstructed tubes are always negative. 
This means that the tubes are always reconstructed closer to the detector. 
It is not clear why this happened, because the number of markers used in the calibration box is the same for both the beamlines, and also the algorithm to evaluate the spatial position is the same.
Further studies will analyze and explain these differences.

Concerning the detectors, the same situation happened on the orientation along the ``c'' direction in the detector 2.
This is surely correlated with what happened in tube 2, and the reason is still unknown.
However, the trueness and precision in both the detectors are sub-millimetric.
Also the orientations are very accurate and do not show any remarkable difference.

The advantage of RSA is the double projection.
In this way, the rough accuracy in the source-detector direction of each beamline is compensated by the other and it is theoretically possible to identify the position of an object in space isotropically.
This is evident in the model position (table \ref{tab:model}: position), where the position accuracy is sub-millimetric.
The same was obtained for the orientations, where the accuracy is very high.
The negative sign in the position trueness means that the model was positioned with a tiny bias to the source position.

The marked cells in the tables represents the values that have a significant inaccuracy. 
The red cells represents values that are greater than zero, and the blue cells the values less than zero.
The white cells are the values that include 0 within their interval of precision.
The highest concentration of colored cells is in the detectors.
Looking at the tables, it is clear the high precision of the evaluations.
This fact makes evident inaccuracies in the scene reconstruction.
These inaccuracies are an effect of the protocol that highlighted the limits of the algorithms used to reconstruct the scene.
Moreover there is a correlation between the trueness and the precision. 
The bigger is the trueness, the bigger is the precision.
Without entering in the algorithm details, the computation of the tube positions is very simple and this introduce few possibilities of error propagation.
The tube position, in fact, had only two border line inaccurate cells.
The calculation of the detector positions/orientations had, on the contrary, a more complex structure and this lead to an important error propagation, making evident the error sources.
The internal sources are the approximation errors within the algorithm.
The external sources are the inaccuracies in the calibration box marker center identification.
However the biases evaluated in the positions/orientations can be used as a correction map to reduce systematic errors of the reconstruction algorithms.

It is important to cite the fact that all the presented results of this work represent the best case. 
The evaluated accuracy is the best limit, while the real experimental accuracy is surely rough. 
This is because in a real setup many unknown and unpredictable variables affect the accuracy, such as, the image bluring, the over- or under-exposure of the image.
Because every real experimental measurement has its own accuracy, it is impossible to evaluate it every time. 
For this reason the best value is reported, and in a real RSA examination it is necessary to keep in mind that the real accuracy is unknown.
What it is assessable, in that case is the uncertainty, according to the GUM definition.

The obtained truenesses are better than the accuracies reported by other motion capture systems, such as Vicon (Vicon Inc., Denver, CO, USA) \citep{Eichelberger2016, Merriaux2017} and Kinect (Microsoft Inc., Redmond, WA, USA) \citep{Otte2016, Mortazavi2018}.
The accuracy of these systems is guaranteed by the high number of cameras that increase the Field of View (FoV) and compensate the discussed error in the source-detector direction or, in this case, source-marker-camera direction.
Although these systems could have a great accuracy in terms of scene reconstruction, the effective movement acquisition is affected by several artifacts \citep{Garling2007, Bonnechere2015}.
Concerning the comparison with other RSA evaluations resporte in the literature, the found accuracy values are very close to the static model-based RSA \citep{Seehaus2016, Bojan2015, Bragdon2002}.
This means that the tested software has algorithms with an efficiency comparable to other systems described in the literature \citep{Stilling2012}.

\begin{figure}[h]
	\centering
	\includegraphics[width=8.6cm]{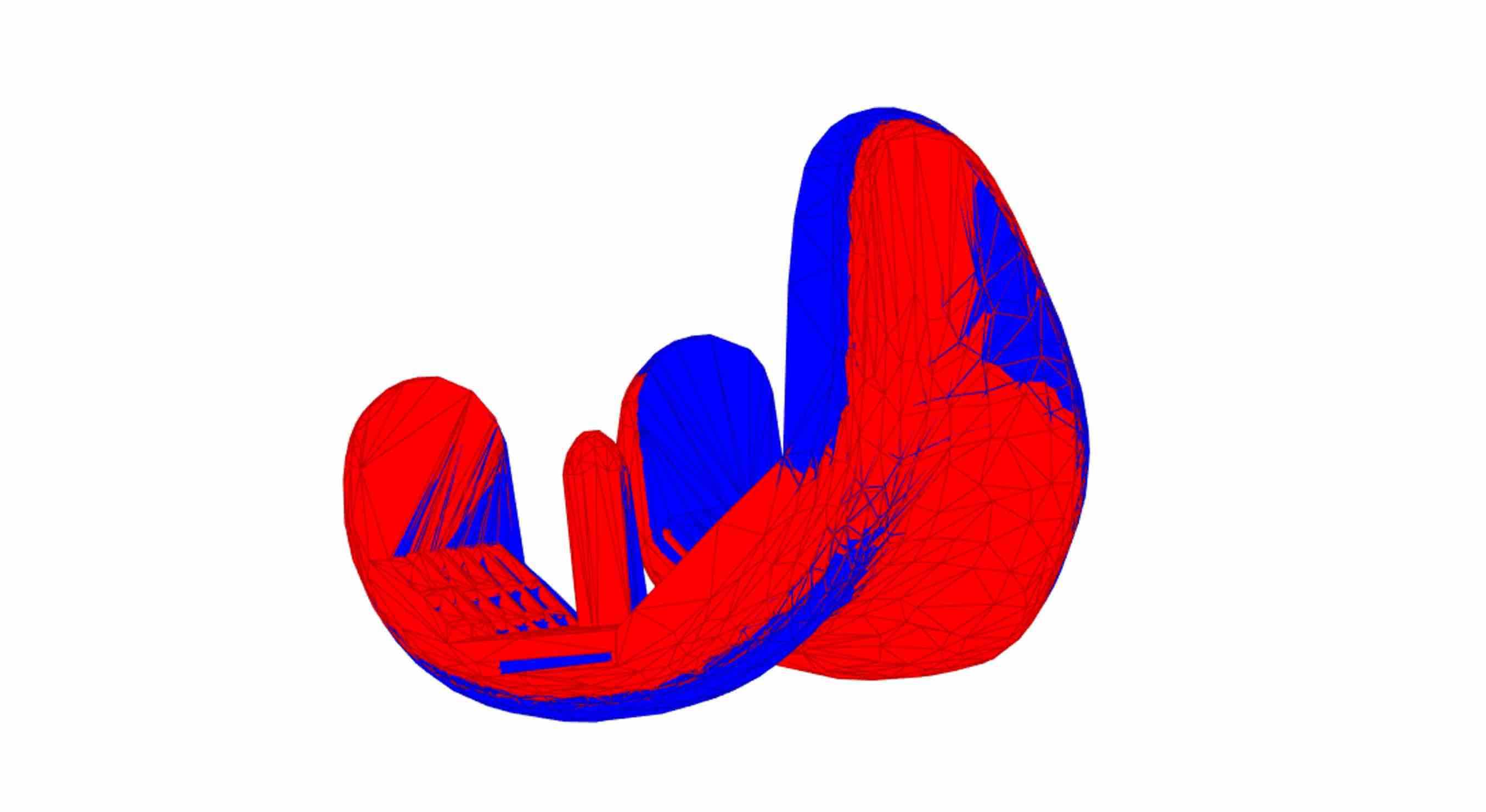}
	\caption{\label{fig:bug} Screenshot of a section of the calibration box that shows the discovered bug in the tested software. A) Original image generated by the simulator (noise level 22.5\%). B) The same image processed by the bugged algorithm. C) The bugged marker detection on the image. D) The image processed with the fixed algorithm. E) The marker detection with the fixed algorithm.}
\end{figure}

During the test on the custom software developed at IOR, some hidden bugs in the scene reconstruction were found and corrected.
These bugs affected the calibration box markers detection (see figure \ref{fig:bug}C). 
The used algorithm was to much sensitive to the level of noise.
In a real x-ray image this bug never happened because of the high signal-to-noise ratio of the image.
In the case of the simulated images, the noise was intentionally structured so the detection algorithm went into crisis (see figure \ref{fig:bug}A and B).
A new preprocessing algorithm was added to the software and the accuracy became independent from the noise level (see figure \ref{fig:bug}D and E).
%~ bug ~~~~~~~~~~~~~~~~~~~~~~~~~~~~~~~~~~~~~~~~~~~~~~~~~~~~~~~~~~~~~~~~~~~~~~~~~~~~~~~~~~~~~
%~~~~~~~~~~~~~~~~~~~~~~~~~~~~~~~~~~~~~~~~~~~~~~~~~~~~~~~~~~~~~~~~~~~~~~~~~~~~~~~~~~~~~~~~~~~

%= CONCLUSION ==============================================================================
\section{Conclusion \label{sec:end}}
The protocol presented in this work was created to be a tool for the evaluation of RSA algorithms. 
It also can be used to compare different softwares.
At last, this simulation system can be used to analyze hidden bugs in algorithms and correct them.
Just like it happened to us during the validation tests of our software.

Another open concern is the different accuracy in the tube reconstruction position and in the detector orientation in the beamline 2.
The answer to this fact deserves a deep analysis of the reconstruction procedure.

This simulator was a first version.
We think that the simulation is the best way to test the data processing codes.
Further improvements will be developed to have more and more realistic simulations, in order to explore setups that are closer to reality and analyze the performances of the algorithms in controlled and specific conditions.

%= ACKNOLEDGEMENTS ========================================================================= 
\section*{Acknowledgements}
We would like to thank Dr. Vittorio Tarabella and Dr. Kevin Ashmore for assistance in revising the language of this manuscript. 

%= CONFLICT OF INTEREST STATEMENT ========================================================== 
\section*{Conflict of interest statement}
The author has no financial or personal relationships with anyone that could inappropriately influence the work presented.

%= BIBLIOGRAPHY ============================================================================
%\section*{References}

%\bibliography{/Users/marco/Documents/Papers/library.bib}

\end{document}